# Hydrothermal activities on C-complex asteroids induced by radioactivity

Wataru Fujiya[1,*], Hisato Higashi[1], Yuki Hibiya[2,†], Shingo Sugawara[1], Akira Yamaguchi[3], Makoto Kimura[1,3] and Ko Hashizume[1]

[1]Faculty of Science, Ibaraki University, 2-1-1 Bunkyo, Mito, Ibaraki 310-8512, Japan

[2]Japan Agency for Marine-Earth Science and Technology, 2-15 Natsushima-cho, Yokosuka, Kanagawa 237-0061, Japan

[3]National Institute of Polar Research, 10-3 Midori-cho, Tachikawa, Tokyo 190-8518, Japan

[*]Corresponding author: wataru.fujiya.sci@vc.ibaraki.ac.jp

[†]Current affiliation: University of Tokyo, Komaba 3-8-1, Meguro-ku, Tokyo 153-0041, Japan

**Abstract**

C-complex asteroids, rich in carbonaceous materials, are potential sources of Earth's volatile inventories. They are spectrally dark resembling primitive carbonaceous meteorites, and thus, C-complex asteroids are thought to be potential parent bodies of carbonaceous meteorites. However, the substantial number of C-complex asteroids exhibits surface spectra with weaker hydroxyl absorption than water-rich carbonaceous meteorites. Rather, they best correspond to meteorites showing evidence for dehydration, commonly attributed to impact heating. Here, we report an old radiometric age of 4564.7 million years ago for Ca-carbonates from the Jbilet Winselwan meteorite analogous to dehydrated C-complex asteroids. The carbonates are enclosed by a high-temperature polymorph of Ca-sulfates, suggesting thermal metamorphism at >300°C subsequently after aqueous alteration. This old age indicates the early onset of aqueous alteration and subsequent thermal metamorphism driven by the decay of short-lived radionuclides rather than impact heating. The breakup of original asteroids internally heated by radioactivity should result in asteroid families predominantly consisting of thermally metamorphosed materials. This explains the common occurrence of dehydrated C-complex asteroids.

1. **Introduction**

Asteroids are small bodies in the solar system. They represent the remnants of planetesimals, i.e., the building blocks of planets. Asteroids are classified by taxonomies based on their surface spectra. C-complex asteroids are characterized by flat or bluish spectra (decreasing reflectance with increasing wavelength) with weak UV features and low visible albedo of ~0.06 on average (Burbine 2014). C-complex asteroids are traditionally linked to primitive, volatile-rich meteorites (carbonaceous chondrites: CCs). The relative abundances and isotopic compositions of H, N, and noble gas suggest the late addition of CC-like materials to the growing Earth, and thus, their nature as potential sources of the Earth's volatiles (Marty 2012; Piani et al. 2020). The late addition of Mighei-type CCs (CM chondrites) by ~0.3% of the Earth's mass can reproduce the isotopic signatures of ruthenium (Ru), a highly siderophile element preferentially partitioned into metallic iron, in the Earth's modern mantle (Fischer-Gödde et al. 2020). Therefore, although small contribution, the corresponding planetesimals are important ingredients of the terrestrial planets especially in terms of their volatile inventories.

The near-infrared spectroscopic survey by the AKARI satellite revealed that 77% and 36% of the observed C-complex asteroids have 2.7 µm and 0.7 µm band features, respectively, which can also be found in the spectra of CM chondrites (Usui et al. 2019). The 2.7 and 0.7 µm band features are attributed to OH in minerals and intervalence charge transfer transition in oxidized iron ($Fe^{2+}$-$Fe^{3+}$), respectively, indicating the presence of phyllosilicates, and thus, past water activities in C-complex asteroids. Nevertheless, the remaining 20-60% of the C-complex asteroids do not exhibit significant band features at such wavelengths. Moreover, the 2.7 µm band depth varies from 4% to 47% among

C-complex asteroids. These observations suggest that the substantial number of C-complex asteroids underwent dehydration to some extent relative to most hydrated CCs. Experimentally or naturally heated/dehydrated CMs exhibit the depth and shape of the 2.7 and 0.7 μm bands well comparable to C-complex asteroids, corroborating the prevailing thermal metamorphism and dehydration processes on C-complex asteroids (Hiroi et al. 1996).

The spatial and temporal scales of thermal metamorphism and dehydration depend on the heat source. Possible heat sources include the decay energy of short-lived radionuclides such as $^{26}$Al (half-life: 0.7 million years), impact-induced heating from the collision with another body, and solar radiation. Impact heating along with rapid cooling has been inferred as a plausible scenario from the nonequilibrated characteristics of thermally metamorphosed CMs comparable to experimentally heated CMs (Nakamura 2005; Nakato et al. 2008). However, the laboratory experiments were conducted in a short time scale of days or weeks, not in a long, geological time scale as long as tens of millions of years (Myr) likely applicable to the heating/cooling by radioactivity. A critical constraint to distinguish between radioactivity and the other heat sources is provided from the chronological information about thermal metamorphism because short-lived radionuclides almost completely decay within a few Myr in the earliest history of the solar system. Thus, if the thermal metamorphism in heated CMs took place within a few Myr after the birth of the solar system, then the plausible heat source is radioactivity. However, such information is currently scarce.

In this work, we studied the Jbilet Winselwan meteorite. We conducted (i) detailed observations to characterize this meteorite in terms of aqueous alteration and thermal metamorphism, (ii) isotopic measurements of Cr and Ti to investigate a genetic link between this meteorite and CM chondrites, (iii) $^{53}$Mn-$^{53}$Cr dating (half-life: 3.7 Myr) of carbonates to obtain the chronological information about aqueous alteration and thermal metamorphism, and finally, (iv) the numerical simulation of the thermal history of the Jbilet Winselwan parent body.

2. Materials and Methods

We observed a polished thin section of Jbilet Winselwan using field-emission scanning electron microscopes (FE-SEM) with energy dispersive spectrometers (EDS) and located characteristic minerals such as Ca carbonates, Ca sulfates, tochilinite/cronstedtite intergrowth (TCI), Fe-Ni sulfides, and Fe-Ni metal. For selected Fe-Ni metal grains, we measured their chemical compositions using an electron probe microanalyzer (EPMA). The polymorphs of Ca carbonates and Ca sulfates were identified with Raman spectroscopy. Subsequently, we performed Mn-Cr isotope analysis of Ca carbonates using secondary ion mass spectrometry (SIMS).

We also prepared powdered Jbilet Winselwan samples to measure its volatile contents and bulk isotopic compositions. We conducted thermogravimetric analysis (TGA) to investigate the volatile loss during heating experiments and estimated the water abundance. We also analyzed the Cr and Ti isotopic compositions using thermal ionization mass spectrometry (TIMS) and multi-collection inductively coupled plasma mass spectrometry (MC-ICP-MS), respectively. The details of the analytical methods are given in Appendix.

3. **Results and Discussion**

We found evidence for aqueous alteration, i.e., the presence of secondary minerals such as carbonates produced under the presence of liquid water (Krot et al. 2015). Calcite ($CaCO_3$), as identified by Raman spectroscopy, is the only carbonate mineral we found. The abundance of Fe-Ni metal is ~0.26 vol%. Jbilet Winselwan contains chondrules with an average diameter of ~0.29 mm ($N$ = 39). The chondrule mesostasis is completely altered whereas the chondrule phenocryst is only partially replaced by phyllosilicate. These petrological characteristics are consistent with the classification of this meteorite as a CM chondrite with petrologic subtypes 2.5-2.3 based on the Rubin et al. (2007) scale, indicating mild aqueous alteration. This classification is also consistent with that by King et al. (2018), who assigned petrologic subtypes 2.7-2.4 to this meteorite. Furthermore, the Cr and Ti isotopic compositions of Jbilet Winselwan (Table A1), in particular, $\varepsilon^{54}Cr$ and $\varepsilon^{50}Ti$ values ($\varepsilon^{54}Cr = 0.86 \pm 0.10$ and $\varepsilon^{50}Ti = 2.80 \pm 0.31$; $N = 5$, where $\varepsilon^{54}Cr$ and $\varepsilon^{50}Ti$ denote parts per 10,000 deviations of $^{54}Cr/^{52}Cr$ and $^{50}Ti/^{47}Ti$ ratios, respectively, from the terrestrial standard values) are identical to those of CMs within the errors (Figure 1). The identical Cr and Ti isotopic compositions indicate a genetic link between this meteorite and CM chondrites.

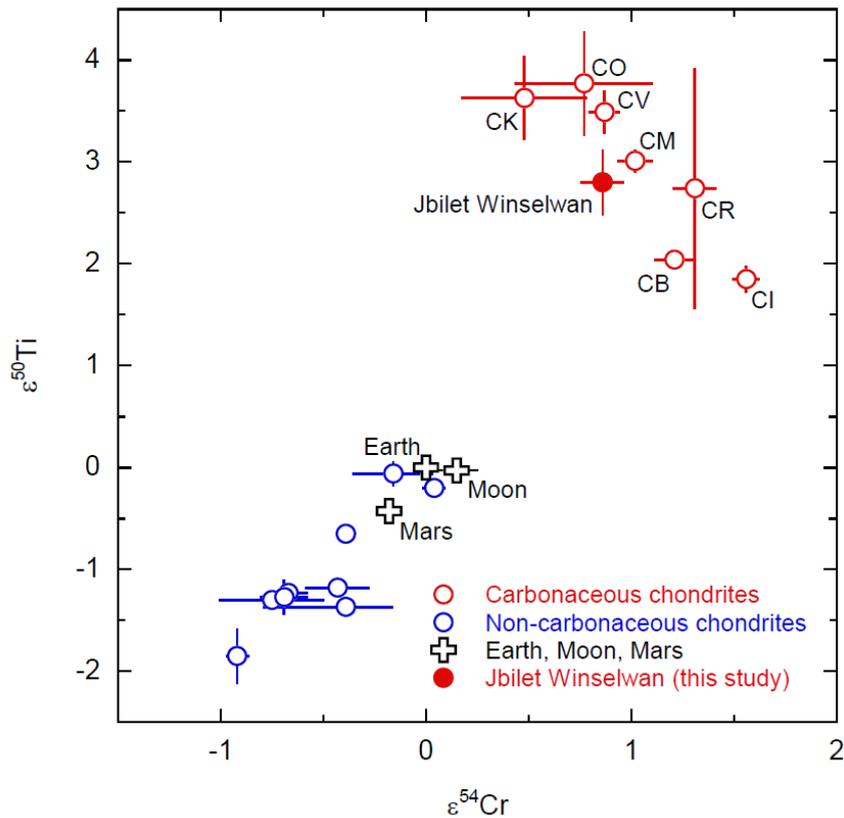

**Figure 1.** Chromium and Ti isotopic compositions of bulk meteorites. $\varepsilon^{54}$Cr and $\varepsilon^{50}$Ti denote parts per 10,000 deviations of $^{54}$Cr/$^{52}$Cr and $^{50}$Ti/$^{47}$Ti ratios, respectively, from terrestrial standard values. The $\varepsilon^{54}$Cr and $\varepsilon^{50}$Ti values of Jbilet Winselwan are identical to those of CM chondrites. Errors on the Jbilet Winselwan data are the 2 standard deviation of the replicated measurements ($N = 5$). Data of the other meteorites are from Burkhardt et al. (2019).

While Jbilet Winselwan is aqueously altered, this meteorite is also thermally metamorphosed. A previous study by King et al. (2018) has reported a lack of coherent X-ray diffraction from the phyllosilicates in this meteorite, which is also commonly observed for heated CMs (Nakamura 2005). In this work, we found that the sulfide minerals in this meteorite are pyrrhotite with pentlandite blebs produced by the decomposition of mono-sulfide solid solution stable at high temperatures of >200°C (Figure A1) (Kelly & Vaughan 1983). This texture is common in thermally metamorphosed CMs (Kimura et al. 2011). The tochilinite ($6Fe_{0.9}S·5(Fe,Mg)(OH)_2$), a mineral commonly found in CMs, in this meteorite is partially decomposed. Tochilinite is stable only at low temperatures of <170°C and likely precipitated at 120°C-160°C (Zolensky 1984; Vacher et al. 2019b). Heating experiments showed that tochilinite decomposes to troilite at ~300°C (Tonui et al. 2014) and the texture of the decomposed tochilinite in Jbilet Winselwan resembles that of an experimentally heated CM chondrite (Nakato et al. 2008; Figure A2). We performed TGA for Jbilet Winselwan twice and found a mass loss of 5.4 and

4.9 wt% between 200°C and 770°C during the heating experiments (Figure A3), which corresponds to the $H_2O$ release from hydrous minerals (Garenne et al. 2014). These values are in good agreement with the water abundance of 4.2 wt% estimated by Vacher et al. (2020) and lower than those of Murchison, an unheated CM, as a reference (9.8 ± 0.8 wt%; 1σ, $N$ = 5). In particular, we observed little mass loss for Jbilet Winselwan between 200°C and 400°C (1.1 and 0.8 wt%), during which hydroxyl of tochilinite is expected to be liberated, consistent with the textural evidence for tochilinite decomposition. Finally, the homogeneous Co and Ni concentrations of Fe-Ni metal indicate the redistribution of these elements during thermal metamorphism (Figure A4; Kimura et al. 2011). Altogether, Jbilet Winselwan underwent aqueous alteration as well as thermal metamorphism, and thus, it is a meteorite analogous to dehydrated C-complex asteroids.

King et al. (2018) estimated a peak metamorphic temperature of 400°C-500°C for Jbilet Winselwan. This lower limit was derived from comparison between the volatile depletion of Jbilet Winselwan and heating experiments conducted in a short timescale. However, a realistic value could be lower than currently estimated if the volatile loss was somewhat related to a diffusional process and the heating timescale was much longer than the experiments. Our SIMS measurements demonstrate that the analyzed calcites have variable Mn contents by a factor of 2 or more, with the $^{55}Mn^+/^{44}Ca^+$ ratios of (1.1-2.6) × $10^{-2}$. The qualitative analyses of the calcites by SEM-EDS also revealed that the Mg, Mn, and Fe abundances vary by a factor of 4-7. Thus, the peak metamorphic temperature of Jbilet Winselwan was likely below the closure temperatures of these elements in calcite so that their abundances were not homogenized by diffusion. The closure temperature of an element in a given mineral is a function of the diffusion coefficient of the element of interest, the cooling rate, and the grain size of the mineral. Assuming the cooling rate of 10°C-100 °C Myr$^{-1}$ and using the observed grain size of 10 μm, the closure temperature of Mg in calcite can be calculated to be ~400°C (Fisler & Cygan 1999). Therefore, the peak temperature of Jbilet Winselwan was likely lower than 400°C if the cooling rate was as slow as 10°C-100 °C Myr$^{-1}$.

The Mn-Cr isotope analysis of the calcites in Jbilet Winselwan showed a linear correlation between the $^{55}Mn/^{52}Cr$ and $^{53}Cr/^{52}Cr$ ratios (Figure 2(A) and Table A2). This correlation likely represents an isochron rather than a mixing line (Figure 2(B)). Thus, the Mn-Cr systematics provides the chronological information about the initial $^{53}Mn/^{55}Mn$ ratio, $(^{53}Mn/^{55}Mn)_0$, at the time of calcite formation. The $(^{53}Mn/^{55}Mn)_0$ ratio of Jbilet Winselwan is (4.51 ± 1.06) × $10^{-6}$ (2σ), corresponding to the formation age of ~4564.7 (+1.1/–1.4) million years ago (Ma) or 2.6 Myr after solar system formation, whereas the carbonates in unheated CMs commonly have significantly younger ages of 4562.7 (+0.4/–0.5) Ma on average (Fujiya et al. 2012; Figure 2(C)). Note that (i) D'Orbigny angrite with a $(^{53}Mn/^{55}Mn)_0$ ratio of 3.54 × $10^{-6}$ (McKibbin et al. 2015) and a Pb-Pb age of 4563.37 Ma (Brennecka & Wadhwa 2012) was used as a time anchor to calculate the age of Jbilet Winselwan, and that (ii) the age of solar system formation, 4567.3 Ma, is recorded by the Pb-Pb ages of the oldest solids in the solar system (Ca- and Al-rich Inclusions, CAIs; Connelly et al. 2012). Recently, Bischoff et al. (2021) reported a similar formation age of 4564.6 ± 1.0 Ma for the carbonates from a unique chondrite Flensburg, which was also heated. All the analyzed calcites are subhedral or anhedral crystals and not polycrystalline, and do not contain Fe-Ni inclusions. Thus, they are 'type 1' or 'T1' calcite and likely precipitated in the earliest stage of aqueous alteration under low temperatures of 0°C -70°C (Guo & Eiler 2007; Lee et al. 2014; Vacher et al. 2019a, 2019b). Jbilet Winselwan is known to be a brecciated CM (King et al. 2018) and its petrology varies among specimens and/or individual clasts. Nevertheless, since the petrological characteristics of the calcites analyzed here are similar, we argue that they formed under similar physicochemical conditions and at almost the same time. The mean square weighted deviation value of the isochron diagram (1.06; Figure 2(A)) corroborates this argument and there is no evidence that the individual calcite grains formed during distinct alteration events. Therefore, it is likely that the carbonate ages approximately represent the timing of the onset of aqueous alteration when $H_2O$ ice in the CM parent body melted and liquid water was available.

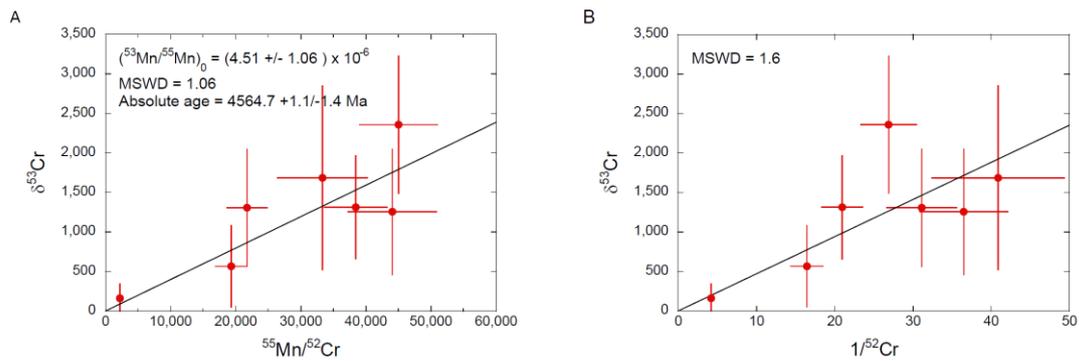

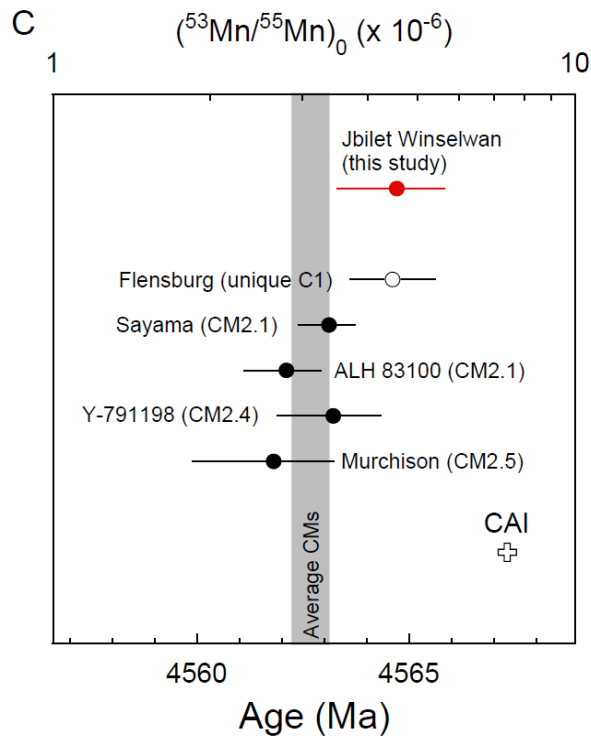

**Figure 2.** Results of the $^{53}$Mn-$^{53}$Cr dating of Jbilet Winselwan calcites. (A) Isochron diagram showing $^{55}$Mn/$^{52}$Cr ratios vs. $\delta^{53}$Cr values ($\delta^{53}$Cr denotes permil deviation of $^{53}$Cr/$^{52}$Cr ratios from the terrestrial standard value). The individual data points show a linear correlation, corresponding to the initial $^{53}$Mn/$^{55}$Mn ratio, ($^{53}$Mn/$^{55}$Mn)$_0$, of (4.51 ± 1.06) × 10$^{-6}$. Note that the intercept of the y-axis is fixed to zero. (B) Chromium contamination plot showing 1/$^{52}$Cr vs. $\delta^{53}$Cr values. The observed correlation appears worse than that in the isochron diagram as shown by their mean squared weighted deviation (MSWD) values. Thus, it is likely that the linear correlation in the isochron diagram provides chronological information and is not produced by the mixing between a $^{53}$Cr-rich component and contamination with the terrestrial Cr isotopic composition ($\delta^{53}$Cr = 0) without any chronological information. (C) Comparison between the formation ages of Jbilet Winselwan and unheated CM chondrites (Fujiya et al. 2012). The age of the Flensburg unique C1 chondrite is also shown (Bischoff et al. 2021). The gray band shows the weighted average of the unheated CM chondrites with the 2σ error. CAI shows the age of solar system formation (4567.3 Ma).

Jbilet Winselwan contains not only Ca carbonates but also Ca sulfates (Figure 3(A)), a rare mineral in CMs (Lee 1993; Brearley 1995). It appears from their texture that the Ca sulfates did not replace other minerals but precipitated from aqueous fluids. Raman spectroscopy revealed their polymorph as anhydrite (CaSO$_4$) stable at temperatures higher than 300°C (Prieto-Taboada et al. 2014), providing another line of evidence for thermal metamorphism (Figure 3(B)). Gypsum (CaSO$_4$·2H$_2$O), stable at lower temperatures, was not found, although King et al. (2018) reported the presence of both anhydrite and gypsum. The sulfur in the Ca sulfate was likely supplied from tochilinite during thermal metamorphism because tochilinite is unstable and decomposed at high temperatures of 300°C. This indicates that the anhydrites precipitated from high-temperature fluids containing dissolved sulfur from tochilinite. Note that Brearley (1995) also reported the presence of anhydrite in Bells, an unusual CM, and suggested the oxidation of sulfides to form $SO_3$ species. The anhydrites in Jbilet Winselwan commonly enclose calcites indicating successive formation of calcite followed by anhydrite, i.e., thermal metamorphism occurred subsequently after aqueous alteration. Given the calcite age measured here, the thermal metamorphism of Jbilet Winselwan took place within a few Myr after the birth of the solar system. Therefore, the most plausible heat source for thermal metamorphism is the decay energy of short-lived radionuclides represented by $^{26}Al$. The older carbonate age of Jbilet Winselwan than unheated CMs attests this conclusion because a larger $^{26}Al$ amount at the onset of aqueous alteration must have resulted in higher peak temperatures for Jbilet Winselwan than for unheated CMs. The recently reported, young $^{87}Rb$-$^{87}Sr$ ages of heated CMs (>3 billion years after the birth of the solar system) likely date the collisional events which destroyed the original asteroids (Amsellem et al. 2020), but do not represent the ancient thermal metamorphism induced by $^{26}Al$ decay.

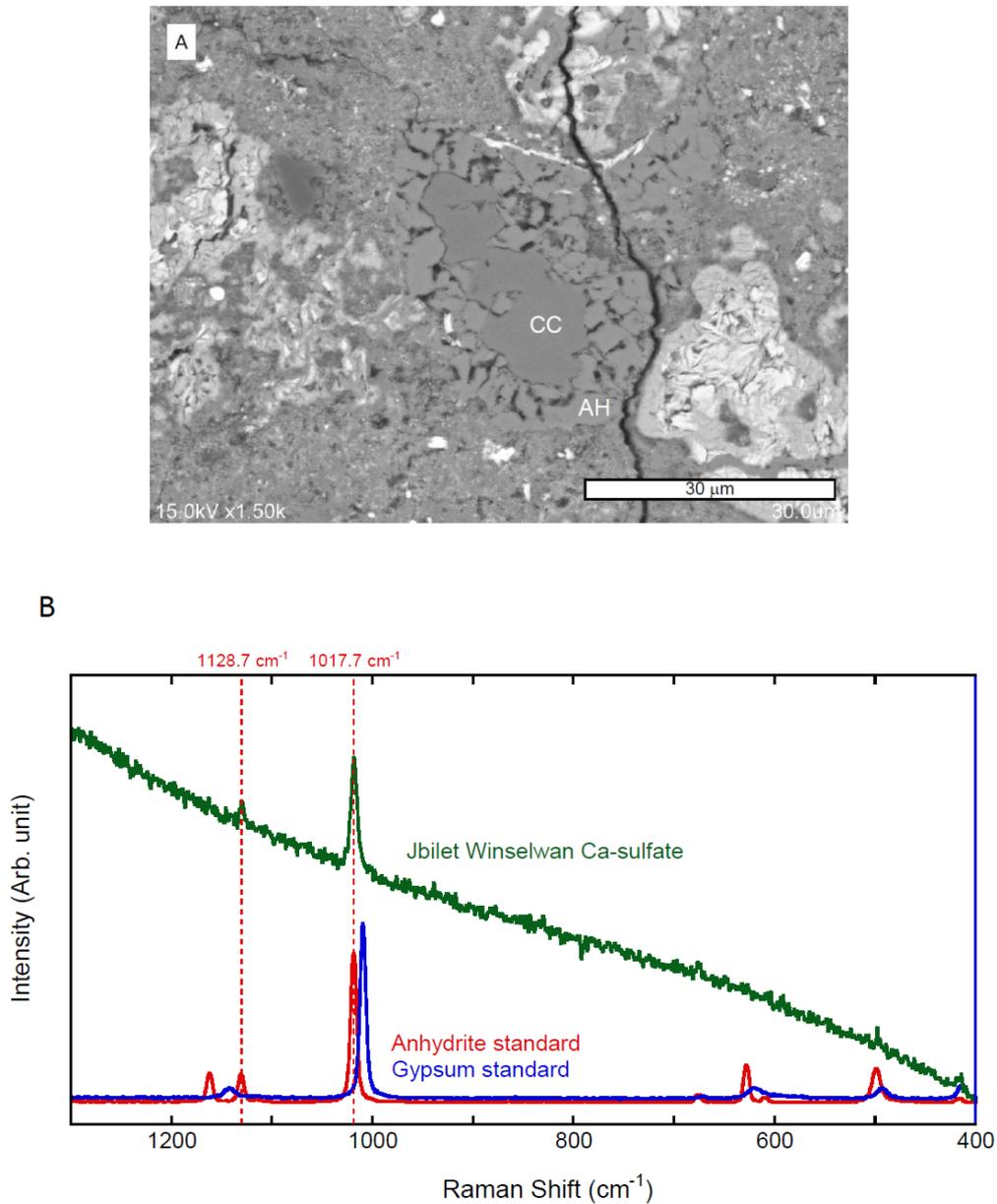

**Figure 3.** Ca sulfate in Jbilet Winselwan. (A) Backscattered electron image of a calcite (CC) grain surrounded by Ca sulfate (AH). This is a typical occurrence of Ca sulfate throughout the matrix of this meteorite, suggesting that the anhydrite precipitation by hydrothermal alteration at >300°C took place subsequently after the calcite precipitation by low-temperature (0°C-70°C) aqueous alteration. (B) Raman spectrum of the Ca sulfate. The peaks of the Raman shift observed for Jbilet Winselwan are consistent with those of anhydrite showing the characteristic Raman bands at 1017 and 1129 cm$^{-1}$ (Prieto-Taboada et al. 2014)

The relationship between Jbilet Winselwan and unheated CMs is unclear. They could be derived from separate parent bodies, however, the similar Cr and Ti isotopic compositions favor similar source reservoirs for both meteorites. If they originate from a single parent body, then Jbilet Winselwan, and possibly, other thermally metamorphosed CMs must have been located in the central part of the parent body relative to unheated CMs.

To investigate the relationship between heated CMs and unheated CMs, especially in terms of the peak temperatures and the timing of the onset of aqueous alteration, we conducted a numerical simulation on the thermal history of the CM chondrite parent body (planetesimal) following a previous study by Fujiya et al. (2012). We assumed a spherically symmetric planetesimal of 50 km in radius and numerically solved a thermal conduction equation with $^{26}$Al decay as a heat source. The ambient temperature was fixed to 170 K. The solid-liquid phase transition of $H_2O$ and the reaction heat of the phyllosilicate formation from anhydrous silicates are also incorporated. The formation time of the planetesimal determines the initial abundance of $^{26}$Al. The planetesimal is assumed to have formed 2.8 Myr after the birth of the solar system, which is 0.7 Myr earlier than estimated by Fujiya et al. (2012). This formation time is later than the apparent formation age of the Jbilet Winselwan calcites (2.6 Myr after solar system formation), however, the timing of the planetesimal formation was determined so that the expected peak temperature is consistent with the observation of Jbilet Winselwan.

The results of the calculation showed that the temperatures in the central part reach ~350°C, consistent with the inferred peak temperature of Jbilet Winselwan. The peak temperatures at the depth of ~10 km and ~20 km are 200°C and 300°C, respectively, and thus, roughly 20 vol% of the planetesimal is heated to 300°C. While the onset of aqueous alteration in the inner regions is ~0.3 Myr after the planetesimal formation, it delays in the outer regions (within 5 km from the surface) by 0.3-1 Myr because the heat from $^{26}$Al decay balances with the heat conduction to the surface (Figure 4). This delay may explain the observed time difference between the carbonate ages of Jbilet Winselwan and unheated CMs. The early onset of aqueous alteration in the inner regions (0.3 Myr after the planetesimal formation or 3.1 Myr after CAIs) is consistent with the old formation age of Jbilet Winselwan (2.6 +1.4/–1.1 Myr after CAIs), while the later onset of aqueous alteration in the outer regions (0.6-1.3 Myr after planetesimal formation or 3.4-4.1 Myr after CAIs) corresponds to the younger formation ages of the carbonates in unheated CMs (4.6 +0.5/–0.4 Myr after CAIs). Note that the carbonate ages likely record the timing of the onset of aqueous alteration or shortly after that.

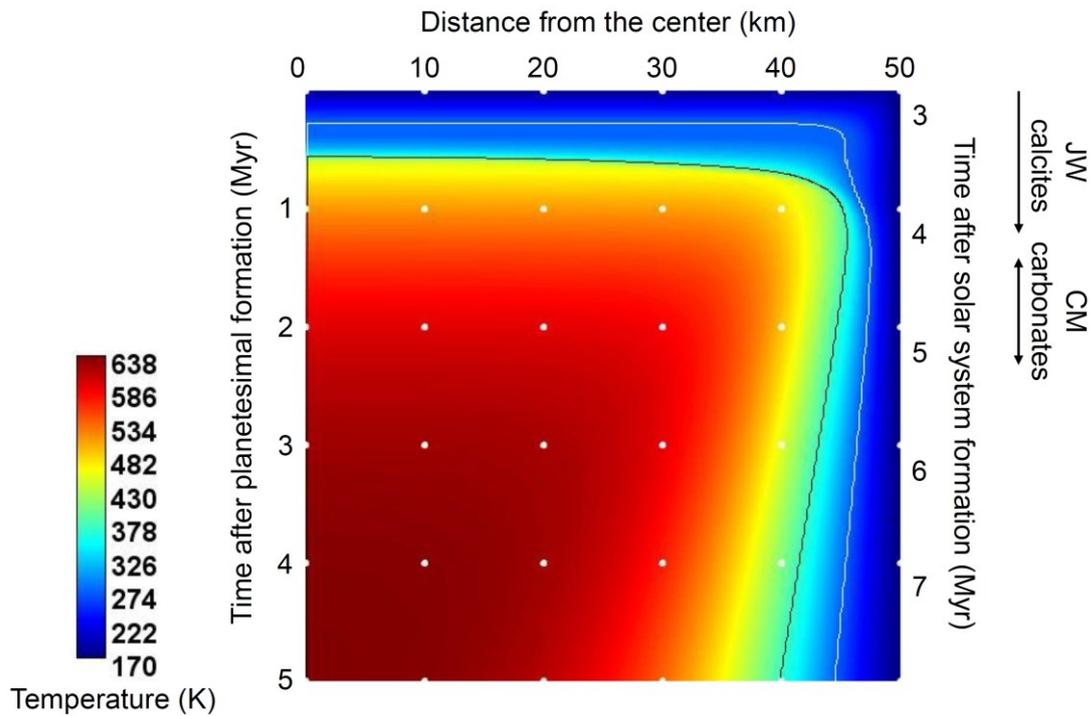

**Figure 4.** Thermal history of the Jbilet Winselwan parent body. The white and black lines show the contour lines of 273 K (0°C) and 343 K (70°C), respectively, representing the range of the calcite formation temperatures. The white dots show 10 km (horizontal) and 1 Myr (vertical) intervals. The formation ages of Jbilet Winselwan (JW) calcites (this study) and CM carbonates (Fujiya et al. 2012) are shown by arrows. The white line demonstrates that the aqueous alteration in the inner regions initiated ~0.3 Myr after the planetesimal formation, while the aqueous alteration in the outer regions delays by up to 1 Myr. CM calcites likely precipitated in the earliest stage of aqueous alteration under low temperatures of 0°C-70°C (between white and black lines). Thus, Jbilet Winselwan may be derived from the inner regions of the parent body while unheated CMs chondrites can also be produced in the outer regions of the same parent body.

Our Mn-Cr data and the numerical simulation demonstrate that at least some thermally metamorphosed CMs like Jbilet Winselwan and unheated CMs can share the same, stratified parent body with their chronological constraints from carbonates being satisfied. The parent body likely formed earlier to achieve higher peak temperatures than the previous estimate of ~3.5 Myr after the birth of the solar system (Fujiya et al. 2012). We note, however, that some CMs are apparently heated to higher temperatures (ca. 750°C) than the peak temperature of Jbilet Winselwan (Nakamura 2005), and that these heated CMs may not be derived from the same parent body as that of Jbilet Winselwan and unheated CMs. If these CMs also originate from the same parent body, then they were likely heated by impact. Thus, we do not argue that all the CMs originate from the same parent body, or that all the thermally metamorphosed CMs were heated by radioactivity. Our calculation also predicts that asteroid families resulting from the breakup of original, larger asteroids like the Jbilet Winselwan parent body predominantly consist of thermally metamorphosed materials. This explains the common occurrence of dehydrated C-complex asteroids.

**Conclusions**

In this work, we studied the Jbilet Winselwan meteorite and found that this meteorite is a CM chondrite with the evidence for aqueous alteration as well as thermal metamorphism. The $^{53}$Mn-$^{53}$Cr age of the Jbilet Winselwan carbonates is older than those of unheated CMs. The old age indicates the earlier onset of aqueous alteration in Jbilet Winselwan than in unheated CMs, when $^{26}$Al was more abundant enough to induce subsequent thermal metamorphism at 300°C-400°C. The numerical simulation of the thermal history of the CM parent body demonstrates that thermally metamorphosed CMs like Jbilet Winselwan and unheated CMs could be derived from the same parent body, as inferred from their similar Cr and Ti isotopic compositions, where thermally metamorphosed CMs were located in the central part. This study suggests that the dehydration accompanied by thermal metamorphism in (at least some) C-complex asteroids were induced by $^{26}$Al decay, i.e., not by surface processes like solar radiation or the impact of other small bodies, but by an internal process that took place in the original, larger asteroids.

*Acknowledgements* – We thank Miyuki Takeuchi for the technical support on the NanoSIMS. We thank an anonymous reviewer for constructive comments and Maria Womack for editorial handling. This work was supported by JSPS KAKENHI grant Numbers 17H02992 and 18H04454, by NIPR Research Project Fund KP307, by the Astrobiology Center Program of National Institutes of Natural Sciences (NINS) (grant No. AB032004), and by NIMS microstructural characterization platform as a program of "Nanotechnology Platform" of the Ministry of Education, Culture, Sports, Science and Technology (MEXT), Japan, grant No. JPMXP09A19UT0157.

**Appendix**

**Analytical Procedures**

We observed a polished thin section of Jbilet Winselwan using field-emission scanning electron microscopes (FE-SEM: JEOL JSM-7100F at National Institute of Polar Research, NIPR, and Hitachi High-Tech S-4800 at Ibaraki University). We measured the chemical compositions of selected Fe-Ni metal grains in Jbilet Winselwan using an electron probe microanalyzer (EPMA: JEOL JXA-8200 at NIPR). We measured Si, Mg, Ni, S, Fe, Cr, P, and Co concentrations using a 15 keV/10 nA electron beam of 5 μm in diameter and ZAF corrections. The standards used in this work were olivine (Si and Mg), metallic nickel (Ni), iron sulfide (S), metallic iron (Fe), metallic chromium (Cr), iron phosphide (P), and metallic cobalt (Co). The detection limits are 32 ppm for Si, 35 ppm for Mg, 140 ppm for Ni, 58 ppm for S, 170 ppm for Fe, 94 ppm for Cr, 95 ppm for P, and 180 ppm for Co. The interference of Fe K$\beta$ on Co K$\alpha$ was mathematically corrected. The polymorphs of Ca carbonate and Ca sulfate were identified with Raman spectroscopy (JASCO NRS-1000 at NIPR). We used a 532 nm laser with a laser power of 2.0-2.2 mW, and the measurement time was 5-30 seconds.

We analyzed the Mn-Cr systematics of calcites in Jbilet Winselwan using secondary ion mass spectrometry (SIMS: CAMECA NanoSIMS 50L at University of Tokyo). We scanned 3 × 3 μm$^2$ areas on the sample surface using an O$^-$ primary ion beam of 150 pA (~1-2 μm in size) and detected secondary ions of $^{44}$Ca$^+$, $^{52}$Cr$^+$, $^{53}$Cr$^+$, and $^{55}$Mn$^+$ simultaneously with four secondary electron multipliers. The count rate of $^{44}$Ca$^+$ was ~4 × 10$^4$ cps. For the calibration of $^{55}$Mn/$^{52}$Cr ratios of the Jbilet Winselwan calcites, we used a synthetic calcite standard doped with Mn and Cr (Sugiura et al. 2010). After the SIMS measurements, we measured the Mn/Cr ratios of the synthetic calcite standard using the same EPMA as that for the Fe-Ni metal. The observed Mn/Cr relative sensitivity factor, defined by the ratio of $^{55}$Mn$^+$/$^{52}$Cr$^+$ ion intensity ratios measured with the SIMS to $^{55}$Mn/$^{52}$Cr ratios measured with the EPMA, was 0.75 ± 0.03 (2$\sigma$), consistent with the value reported by Sugiura et al. (2010).

We measured the volatile contents with thermogravimetric analysis (TGA: Rigaku TG-DTA8122 at Ibaraki University). Sample aliquots of Jbilet Winselwan as well as the Murchison CM2 chondrite, approximately 10 mg, were put in Pt pans and analyzed under an N$_2$ flow of 100 ml min$^{-1}$. We recorded the mass loss of the samples by increasing temperature from room temperature to 1000°C. To check the reproducibility of the measurements, we repeated the measurements twice for Jbilet Winselwan and five times for Murchison. The stability of the instrument is ±60 μg, i.e., ~0.6 wt% of the samples. From the five times measurements of Murchison, we obtained a mass loss of 9.8 ± 0.8 wt% (1$\sigma$) between 200°C and 770°C corresponding to the water loss from hydrous minerals, i.e., tochilinite and phyllosilicates (Garenne et al. 2014).

We also measured the Cr and Ti isotopic compositions of Jbilet Winselwan using thermal ionization mass spectrometry (TIMS: Thermo Fisher Scientific Triton Plus at JAMSTEC) and multi-collection inductively coupled plasma mass spectrometry (MC-ICP-MS: Thermo Fisher Scientific, Neptune Plus at JAMSTEC), respectively. Chromium isotopic measurements were conducted following the method described by Yamakawa et al. (2009). Sample aliquots of approximately 40 mg were weighed into clean 3ml Teflon® vials and digested with 1.2 ml concentrated HF and 0.9 ml concentrated $HNO_3$ in a Parr bomb at 200°C for >65 hr. The digested samples were converted to soluble compounds by repeated evaporation with concentrated $HNO_3$ and finally re-dissolved in 1 ml 6M HCl on a 140°C hotplate. Both Cr and Ti were separated from the digested samples following the procedure described by Hibiya et al. (2019), using Bio-Rad AG1-X8 anion exchange resin (1 ml; 200-400 mesh), Eichrom pre-packed 2 ml cartridges containing TODGA resin (200-400 mesh), and Bio-Rad AG50W-X8 cation exchange resin (100-200 mesh). All the Cr isotopic ratios are presented as relative deviation from the NIST-979 Cr standard in standard epsilon notation: $\varepsilon^n Cr = [(^nCr/^{52}Cr)_{sample}/(^nCr/^{52}Cr)_{ref} - 1] \times 10^4$. All the Ti isotopic ratios are presented as relative deviation from the NIST-3162a standard in standard epsilon notation: $\varepsilon^n Ti = [(^nTi/^{47}Ti)_{sample}/(^nTi/^{47}Ti)_{ref} - 1] \times 10^4$. The instrumental mass bias was corrected using an exponential law relative to $^{49}Ti/^{47}Ti = 0.749766$ (Leya et al. 2007).

Figure A1 shows a backscattered electron image of sulfide minerals in Jbilet Winselwan. Figure A2 compares backscattered electron images of tochilinite-cronstedtite intergrowth (TCI) in Jbilet Winselwan and the Murchison CM2 chondrite. Figure A3 contrasts the results of TGA on Jbilet Winselwan and Murchison. Figure A4 shows the Ni and Co concentrations of Fe-Ni metal grains in Jbilet Winselwan.

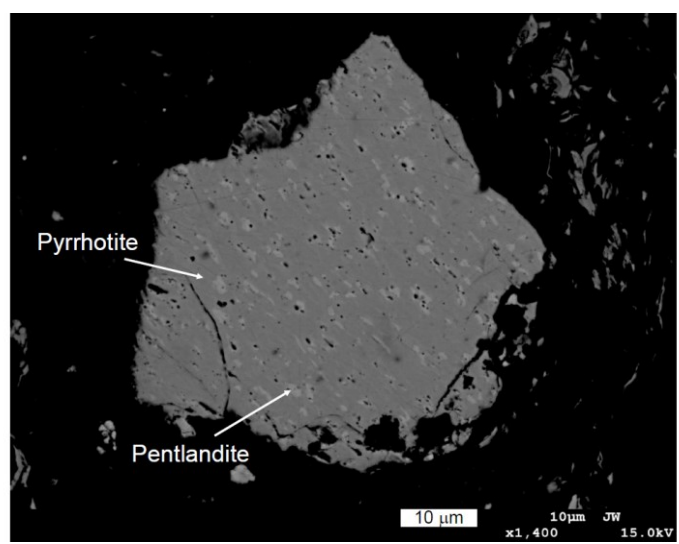

**Figure A1.** Backscattered electron image of sulfide minerals in Jbilet Winselwan. They are pyrrhotite with pentlandite blebs, suggesting thermal metamorphism.

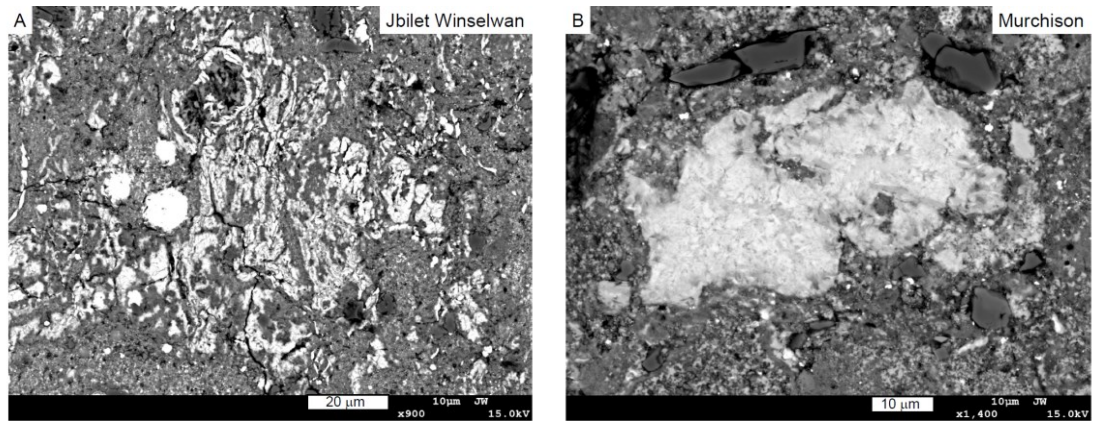

**Figure A2.** Backscattered electron images of tochilinite-cronstedtite intergrowth (TCI) in Jbilet Winselwan (A) and the Murchison CM2 chondrite (B). The TCI in Jbilet Winselwan is partially decomposed.

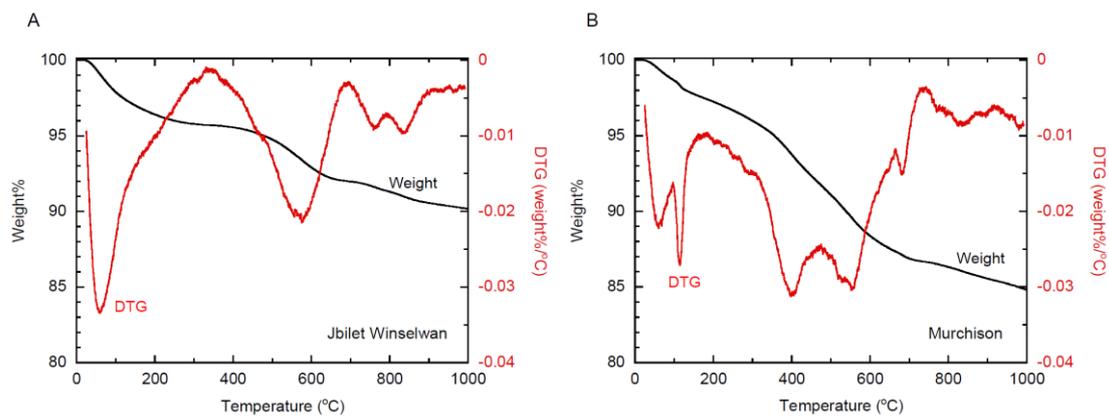

**Figure A3.** Mass loss (black) and its first derivative (DTG: red) curves of Jbilet Winselwan (A) and Murchison (B). The DTG curves can be used to identify the host minerals of the observed volatile loss. Based on the DTG curves of CMs, Garenne et al. (2014) divided the entire volatile loss into four parts: the first part (0°C-200°C) for absorbed and mesopore $H_2O$, the second part (200°C-400 °C) for $H_2O$ from hydroxides, the third part (400°C -770°C) for $H_2O$ from phyllosilicates, and the last part (770°C-900 °C) for $CO_2$ from carbonates with possible contributions from sulfates. Jbilet Winselwan shows a smaller volatile loss than Murchison in total, and in particular, between 200°C and 400 °C, suggesting the volatile loss on its parent body due to thermal metamorphism.

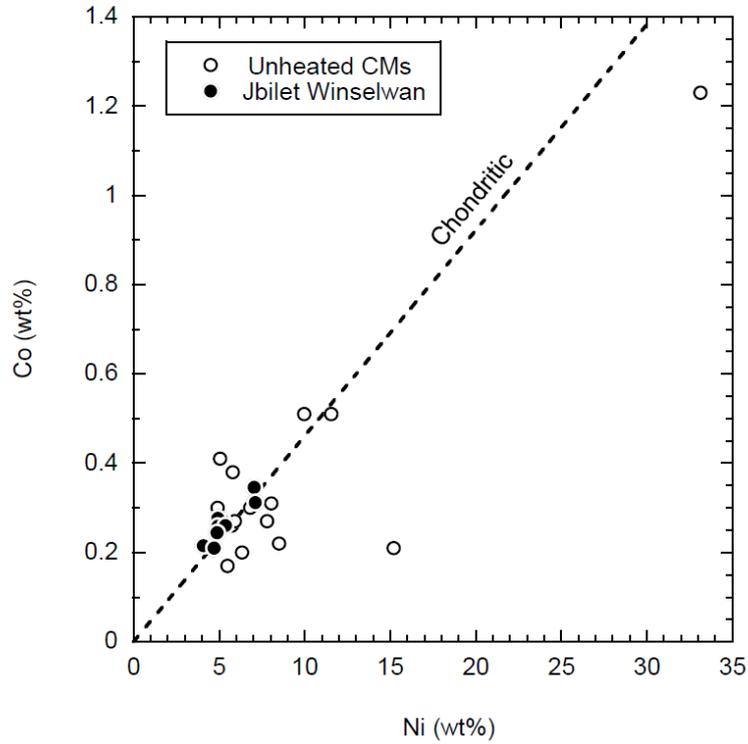

**Figure A4.** Nickel and Co concentrations of Fe-Ni metal grains in Jbilet Winselwan and unheated CMs. The data of Jbilet Winselwan are much less scattered than those of unheated CMs, suggesting homogenization during thermal metamorphism. Data of unheated CMs are from Kimura et al. (2011).

**Table A1**

Chromium and Ti isotopic compositions of the Jbilet Winselwan meteorite

| Sample | $\varepsilon^{53}Cr$ | $2\sigma$ | $\varepsilon^{54}Cr$ | $2\sigma$ | $\varepsilon^{46}Ti$ | $2\sigma$ | $\varepsilon^{48}Ti$ | $2\sigma$ | $\varepsilon^{50}Ti$ |
|---|---|---|---|---|---|---|---|---|---|
| Mean Jbilet Winselwan | 0.16 | 0.04 | 0.86 | 0.10 | 0.40 | 0.18 | -0.11 | 0.07 | 2.80 |
| JW-1 | 0.13 | 0.05 | 0.83 | 0.09 | 0.50 | 0.19 | -0.07 | 0.12 | 2.67 |
| JW-2 | 0.15 | 0.04 | 0.81 | 0.08 | 0.45 | 0.23 | -0.11 | 0.18 | 2.98 |
| JW-3 | 0.15 | 0.07 | 0.84 | 0.13 | 0.43 | 0.17 | -0.09 | 0.11 | 2.72 |
| JW-4 | 0.18 | 0.05 | 0.92 | 0.08 | 0.30 | 0.22 | -0.09 | 0.17 | 2.69 |
| JW-5 | 0.17 | 0.07 | 0.91 | 0.08 | 0.31 | 0.24 | -0.16 | 0.16 | 2.95 |

**Table A2**

Mn- and Cr-isotope data of Jbilet Winselwan calcites

| Grain | $^{55}$Mn/$^{52}$Cr | 2σ | δ$^{53}$Cr (‰) | 2σ | 1/$^{52}$Cr (1/cps) | 2σ |
|---|---|---|---|---|---|---|
| #1 | 38454 | 4809 | 1313 | 652 | 21.0 | 2.6 |
| #2 | 2208 | 116 | 161 | 184 | 4.2 | 0.2 |
| #3 | 44070 | 6804 | 1256 | 793 | 36.5 | 5.6 |
| #4 | 21742 | 3100 | 1307 | 742 | 31.1 | 4.4 |
| #5 | 45017 | 5965 | 2358 | 868 | 26.9 | 3.6 |
| #6 | 33337 | 6892 | 1685 | 1163 | 40.9 | 8.5 |
| #7 | 19323 | 2438 | 566 | 516 | 16.5 | 2.1 |